\documentstyle[aps,prl,multicol]{revtex}

\begin{document}
\title{Inconsistencies of the Adiabatic Theorem and the Berry Phase}

\author{Arun K. Pati$^{(1)}$ and A. K. Rajagopal$^{(2)}$ }
\address{$^{(1)}$ Institute of Physics, Sainik School Post, 
Bhubaneswar-751005, Orissa, India\\}
\address{$^{(2)}$ Naval Research Laboratory, 
Washington D.C. 20375-5320, USA}

\date{\today}
\maketitle
\def\ra{\rangle}
\def\la{\langle}
\def\ver{\arrowvert}

\begin{abstract}
The adiabatic theorem states that if we prepare a quantum system in one of
the instantaneous eigenstates then the quantum number is an adiabatic invariant
and the state at a later time is equivalent to the instantaneous eigenstate
at that time apart from phase factors.
Recently, Marzlin and Sanders have pointed out that this
could lead to apparent violation of unitarity. We resolve the Marzlin-Sanders 
inconsistency within the quantum adiabatic theorem. Yet, our resolution points
to another inconsistency, namely, that the cyclic as well as non-cyclic 
adiabatic Berry phases may vanish under strict adiabatic condition. 
We resolve this inconsistency and develop an unitary operator decomposition 
method to argue for the validity of the adiabatic approximation.

\end{abstract}

\vskip .5cm

PACS           NO:    03.65.-w, 03.65 Ca, 03.65.Vf\\


\begin{multicols}{2}

\vskip 1cm
\section{ Introduction } 
Adiabatic theorem is one of the  most important and 
widely studied theorems in quantum mechanics \cite{lis,mes}. 
It states that if we have a
slowly changing Hamiltonian that depends on time, and the system is
prepared in one of the instantaneous eigenstates of the Hamiltonian
then the state of the  system at any time is given by an
instantaneous eigenfunction of the Hamiltonian up to  multiplicative
phase factors \cite{pe,born,js}. It has potential application in diverse
areas of physics such as in molecular physics, nuclear physics, chemical
physics, quantum field theory and so on. It is the revisit of 
adiabatic theorem that enabled Berry to discover his now famous geometric
phase \cite{berry} which has numerous applications ranging from quantum 
Hall-effect \cite{sw} to understanding of damping of collective excitations
in finite fermi systems \cite{jain}. Furthermore, adiabatic theorem and Berry 
phase have
played important roles in design of quantum algorithms \cite{farhi} 
and implementation of geometric quantum computations \cite{jvec}.

Recently, Marzlin and Sanders (MS) 
have pointed out an inconsistency in the quantum adiabatic
theorem \cite{ms}. In this paper we resolve their inconsistency. However,
our resolution leads to another inconsistency, namely, under strict
adiabatic approximation cyclic as well as non-cyclic Berry phase almost
vanish! We suggest that while dealing with adiabatically changing Hamiltonians
one should not be within the strict regime nor one should be too much away
from adiabatic regime. One has to optimize the operational scale so as
not to encounter MS type inconsistency or vanishing adiabatic Berry phase
type inconsistency. The organization of our paper is as follows. In 
section II, we discuss the Marzlin-Sanders inconsistency. In section III, we 
give a simple resolution to it. In section IV, we point out an apparant 
inconsistency with the cyclic and non-cyclic Berry phase and provide a 
resolution to it. In section V, we develop a unitary operator decomposition 
which can be split into a diagonal and off-diagonal part that captures 
the adiabatic theorem, Berry phase and transition to non-adiabatic regime 
in an elegant way.
Also, we derive a lower bound for the smallness parameter that plays a 
crucial role in validity of the adiabatic theorem. In section VI, we 
provide two examples to illustrate our novel results. Finally our 
conclusion follows.

\section{ The Marzlin-Sanders Inconsistency}  

Consider a quantum system that is governed by a Hamiltonian $H(t)$
with discrete, non-degenerate spectrum $\{ E_n(t) \}$ 
($n=0,1,2,\ldots N $). The instantaneous eigenstates $|n(t)\ra$ of the
Hamiltonian satisfies an  eigenvalue equation
$H(t)|n(t)\ra= E_n(t)|n(t)\ra$.
If the state of the system was initially in one of the eigenstate
(say $n$th), then under adiabatic approximation (AA) the state at a later
time  $t$ is given by
\begin{equation}
|\psi(t)\ra \approx e^{i\delta_n(t) } e^{i \gamma_n(t)} |n(t) \ra
\end{equation} 
where $\delta_n(t) = - \int E_n(t') dt' (\hbar =1)$ 
is the dynamical phase and $\gamma_n(t) =
\int i \la n(t')|{\dot n}(t')\ra dt' $ is the extra phase that is usually
neglected in the text book but can 
give rise to Berry phase during a {\em cyclic change of the
Hamiltonian} \cite{berry}. (Some authors mention that $\gamma_n(t)$ is 
the Berry phase which is incorrect. Unless we have a cyclic evolution 
$\gamma_n(t)$ is not gauage invariant, hence not observable.) 

Since the MS inconsistency is a very curious one, (as well as for the sake of
completeness), we spell out their argument. 
Let us define a state
$|{\bar \psi}(t)\ra = U^{\dagger}(t) |\psi(0)\ra$, where $U(t) = 
T \exp(-i \int H(t)~ dt)$. This state
satisfies a Schr{\"o}dinger equation with a Hamiltonian ${\bar H}(t)$,
i.e.,
\begin{equation}
i |{\dot {\bar \psi}}(t)\ra = {\bar H}(t) |{\bar \psi}(t)\ra,
\end{equation} 
where ${\bar H} = -  U^{\dagger}(t)H(t) U(t)$. 
Note that 
This holds irrespective of adiabatic approximation.
Now, what is the solution to the above equation in the adiabatic
approximation?  
First, note that if $|n(t)\ra$ is an eigenstate of
$H(t)$ with eigenvalue  $E_n(t)$, then $|{\bar n}(t)\ra =
U^{\dagger}(t) |n(t) \ra$ is an eigenstate  of ${\bar H}(t)$ with
eigenvalue $-E_n(t)$. Since $|{\bar n}(t)\ra$'s are  orthonormal and
satisfies completeness criterion, this forms an  instantaneous
basis. We can expand
\begin{equation}
|{\bar \psi}(t)\ra = \sum_n d_n(t) |{\bar n}(t) \ra, 
\end{equation} 
and using AA the amplitudes $d_n(t)$'s are given by
\begin{eqnarray}
id_n(t) + E_n(t) d_n(t) & + & i\la {\bar n}(t)|{\dot {\bar n} }(t) \ra
d_n(t) \nonumber\\
& + & \sum_{m\not=n}  i\la {\bar n}(t)|{\dot {\bar m} }(t) \ra d_m(t) =0.
\end{eqnarray} 
Under adiabatic approximation we can drop the terms like  $\la {\bar
 n}(t)|{\dot {\bar m} }(t) \ra$ because
\begin{equation}
 \frac{ |\la {\bar n}(t)|{\dot {\bar m} }(t) \ra }{|E_n - E_m|} << 1,
~~ n \not=m.
\end{equation}
In general if $H(t)$ varies slowly a unitarily related Hamiltonian
need not vary slowly. But in this case it happens to be so. Under the
above condition the solution is found to be
\begin{equation}
d_n(t) \approx  \exp( -i \delta_n(t) ) \exp(i\int~ i\la {\bar
n}(t)|{\dot {\bar n} }(t) \ra~ dt ) d_n(0).
\end{equation} 
Therefore, the state 
$|{\bar \psi}(t)\ra$ is given by
\begin{equation}
 |{\bar \psi}(t)\ra \approx \exp(- i \delta_n(t) ) \exp( i \int~ i \la
{\bar n}(t)|{\dot {\bar n} }(t) \ra ~dt ) |{\bar n}(t) \ra .
\end{equation} 
Using $|{\bar n}(t)\ra = \exp( - i\delta_n(t) ) \exp(- i \gamma_n(t))
|n(0) \ra$ and $i \la {\bar n}(t)|{\dot {\bar n} }(t) \ra =  i \la
n(t)|{\dot n }(t) \ra - E_n(t)$ one will have
\begin{equation}
|{\bar \psi}(t)\ra \approx e^{i\int E_n(t') dt' } |n(0) \ra.
\end{equation}
This was the result of MS. Now we give their
contradiction. Since $|{\bar \psi}(t)\ra$ is unitarily related
to the initial state, so it must be normalized to unity. However,
the above solution together with the standard adiabatic ansatz (1) 
gives a non-unit norm!
Explicitly, one can see that
\begin{equation}
\la {\bar \psi}(t)| {\bar \psi}(t) \ra = 
\la \psi(0)|U(t)| {\bar \psi}(t) \ra \approx e^{i \gamma_n(t)} 
\la n(0)| n(t) \ra \not= 1  
\end{equation}
Hence a contradiction. So the standard adiabatic theorem apparently violates
unitarity!  Does it really do so?

\section{ Resolution of the MS Inconsistency}  

The resolution to
the above contradiction is now given within the adiabatic theorem. 
Now let us take
a close look at the transition amplitude $A_n(t) = \la n(0)| n(t) \ra$
 between the initial and the instantaneous eigenstates, 
and ask how does that change with time as
we slowly change the Hamiltonian. 
Consider the transition amplitude defined by
\begin{equation}
A_n(t) = \la n(0)| n(t) \ra.
\end{equation}
Its time rate of change is given by ${\dot A}_n(t)$
\begin{eqnarray}
{\dot A}_n(t) =
& = &  
\sum_m  \la n(0)|m(t) \ra \la m(t)| {\dot n(t)} \ra  
= \la n(0)|n(t) \ra \la n(t)| {\dot n(t)} \ra \nonumber\\
&+& \sum_{m\not=n} \la n(0)|m(t) \ra \la m(t)| {\dot n(t)} \ra.
\end{eqnarray}
Now under standard adiabatic approximation one can drop the terms $\la m(t) |
{\dot n(t)} \ra$, for $m\not=n$. Then we have
the transition amplitude $A_n(t)$ as follows:
\begin{eqnarray}
i {\dot A}_n(t) \approx {\dot \gamma_n}(t) A_n(t), 
\end{eqnarray}
where ${\dot \gamma_n}(t)= i \la n(t)| {\dot n(t)} \ra  $ is 
the Berry frequency.
This leads to 
\begin{eqnarray}
 A_n(t) \approx  e^{ - i \gamma_n (t)} A_n(0).
\end{eqnarray}
Since $A_n(0)=1$, this implies that $ \la n(0)| n(t) \ra \approx  e^{ - i
\gamma_n(t)} $. Using this solution one can easily see that
the MS contradiction is resolved, i.e., the unitarity is
preserved.  This also tells us that the transition probability between
initial eigenstate  and the later instantaneous eigenstate is unity
for all time.  In terms of the `minimum-normed distance' \cite{akp} we have
$D^2(|n(0)\ra, |n(t)\ra) =  2(1 - |\la n(0)| n(t) \ra| ) \approx 0$
which is almost zero. So under strict  adiabatic condition the
instantaneous eigenstate apparently stays almost close  to the original
one and hence there is no violation of norm preservation.

\section{ The Berry phase- yet another Inconsistency} 

The above resolution
could have been a satisfying situation if the following is not true. 
However, as we
will show below under strict adiabatic evolution, i.e., under the 
condition (5) the cyclic as
well as the non-cyclic Berry phases almost vanish!

Consider the cyclic variation of the Hamiltonian, i.e., $H({\bf R}(T)) = 
H({\bf R}(0))$ over a period of time $T$. Then we know that the state of 
the system at time $t =T$ is given by
\begin{equation}
|\psi(T)\ra \approx e^{-i\int_0^T E_n(t) dt } e^{i \gamma_n(C)} 
|\psi(0) \ra
\end{equation} 
where $\gamma_n(C)$ is given by 
\begin{equation}
\gamma_n(C) = i \int_0^T \la n(t)| {\dot n(t)} \ra~dt = 
\oint \la n({\bf R})|\nabla n({\bf R}) \ra.d{\bf R}
\end{equation}
is the gauge-invariant Berry phase that depends only on the geometry of the 
path in the parameter space and is also measurable \cite{berry}. 
Therefore, we 
have $\la \psi(0)|\psi(T)\ra = \exp[ i( \delta_n(T) + \gamma_n(C) )] $.
However, our solution
to MS inconsistency suggest that $\la \psi(0)|\psi(T)\ra = \exp(i\delta_n(T))
\exp( i \gamma_n(C))\la n(0)|n(T) \ra = \exp(i\delta_n(T)) $.
This implies that the initial and the final state differ only by the 
dynamical phase and there is no observable Berry phase.
Hence, a contradiction!

Next, consider the general definition of the Berry phase when a pure 
state vector
undergoes a time evolution $|\psi(0)\ra \rightarrow |\psi(t)\ra$. Invoking 
Pancharatnam's idea of relative phase shift one can define the geometric phase 
for non-cyclic evolution of quantum systems \cite{samu}. The
non-cyclic geometric phase is then given by a gauge invariant 
functional of $\psi(t)$ along an open path $\Gamma$ \cite{simon}
\begin{eqnarray}
 \Phi_G[\Gamma] &=& {\rm Arg} \la \psi(0)|\psi(t)\ra + i \int dt \la
\psi(t)|{\dot \psi}(t)\ra 
\end{eqnarray}
Using the reference-section state vector $|\chi(t)\ra$ we can express 
the general geometric phase during an arbitrary evolution as 
\begin{eqnarray}
 \Phi_G[\Gamma] = i \int ~dt \la \chi(t)|{\dot \chi}(t)\ra = \int G,
\end{eqnarray}
where the reference-section $|\chi(t)\ra = \frac{ \la \psi(t)|\psi(0)\ra }
{|\la \psi(t)|\psi(0)\ra|} |\psi(t)\ra$ and $G = 
i \la \chi|d \chi \ra $ is the generalized gauge 
potential or connection form \cite{akp1,akp2}. 
Thus, the generalized geometric phase can be written as a line integral
of a vector potential in the projective Hilbert space of a quantum system.
For differential geometric formulation of the general Berry phase see 
\cite{akp1}.

Note that the above definition holds irrespective of adiabatic,
cyclic,  and Schr{\"o}dinger time evolution. 
So this is the generalized geometric phase during a time evolution 
of a quantum system described by a pure state vector. 
Under adiabatic
approximation there is an open-path Berry phase which was introduced in
\cite{akp3} and is given by
\begin{eqnarray}
 \Phi_G^{(n)}[\Gamma] &=& {\rm Arg} \la n(0)| n(t)\ra + 
i \int ~dt \la n(t)|{\dot
n}(t)\ra \nonumber\\
&=& \int {\bf G}_n({\bf R}). d{\bf R},
\end{eqnarray}
where $\bf G_n({\bf R})$ is the generalized gauge potential 
that gives 
rise to the adiabatic open-path Berry phase. Under strict
adiabatic approximation using our previous solution we find that
$\Phi_G^{(n)}[\Gamma] \approx 0$. Thus, it almost vanishes for all time! 
Therefore,
{\em it appears that under strict adiabatic evolution a quantum system 
cannot acquire any
Berry phase  (non-cyclic as well as cyclic)}. However, there are many
physical systems that show the existence of non-cyclic and cyclic
Berry phases under adiabatic approximation. Thus our solution
, though resolves Marzlin-Sanders inconsistency, yet 
points to another important inconsistency.

Hence, a possible way out is not to drop off-diagonal terms that are
usually done. One has to be careful when to drop and when not
to. The statement that matrix elements causing transition to other
eigenstates are dropped under adiabatic approximation could lead to
internal inconsistencies either MS type or our type
(i.e. the vanishing Berry phase). Thus the source of our inconsistency is
dropping of the terms like $\la m(t) |{\dot n(t)} \ra$, for $m\not=n$.

We have seen that under strict adiabatic approximation the transition
amplitude between the initial and the instantaneous eigenstate obeys a linear
homogeneous equation that results in a vanishing Berry phase. However, if we
investigate the full solution and one can save the adiabatic Berry phase.
Note that Eq(4) can be written as 
\begin{eqnarray}
i \frac{d}{dt} A_n(t) = {\dot \gamma_n}(t) A_n(t) + S_n(t),
\end{eqnarray}
where $S_n(t) = \sum_{m\not=n} \la n(0)|m(t) \ra \la m(t)| {\dot n(t)} \ra$.
The solution to the above equation can be written as
\begin{eqnarray}
A_n(t) = e^{- i \gamma_n(t)} [1 - i \int_0^t~ dt' S_n(t') e^{i \gamma_n(t')} ].
\end{eqnarray}
The second term clearly represents the correction to the adiabatic 
approximation and its presence can only make a non-zero Berry phase
in cyclic as well as non-cyclic case. In particular, the generalized 
non-cyclic adiabatic
Berry phase given by Eq(8) can be expressed as 
\begin{eqnarray}
 \Phi_G^{(n)}[\Gamma] = 
\tan^{-1} \bigl[ \frac{- Re Q_n(t)}{1 + Im Q_n(t) } \bigr],
\end{eqnarray}
where $Q_n(t) = \int~ dt' S_n(t') e^{i \gamma_n(t')}$. This can be shown to 
be related to the response function of a many body quantum system that 
explains damping of collective excitations \cite{jain}.

\section{ Unitary operator for adiabatic evolution} 

We can develop a
general  solution to the unitary evolution operator and show that it
has two  pieces; a diagonal and a non-diagonal piece. It is the
diagonal piece  that gives what we want - adiabatic theorem but there
is no  inconsistency if we keep the order of approximation in our
development.

The time evolution operator of a quantum system obeys
\begin{eqnarray}
 i {\dot U}(t) = H(t) U(t)
\end{eqnarray}
with $U(0) =I$, and the unitarity condition for all $t$ holds, i.e., 
$U(t)U(t)^{\dagger} = U(t)^{\dagger} U(t) = I$, where $I$ is the
idenity  operator. 
Since the Hamiltonian is Hermitian for any $t$, it
admits an  eigen-expansion (for simplicity, we consider here discrete
and non-degenerate  case) $H(t) |n(t)\ra = E_n(t)|n(t)\ra$, $\la
n(t)|m(t) \ra =\delta_{nm}$  and $\sum_n |n(t)\ra \la n(t)| =I$, for all $t$. 
We can express the unitary operator in terms of these
instantaneous eigenstates as 
\begin{eqnarray}
U(t) = \sum_{nm} |n(t)\ra U_{nm}(t) \la m(0|,
\end{eqnarray}
where the matrix elements satisfy
\begin{eqnarray}
i{\dot U}_{nm}(t) + E_n(t) U_{nm}(t) + \sum_{p}  i\la  n(t)|{\dot p }(t)
\ra U_{pm}(t) =0.
\end{eqnarray}
The condition at $t=0$ implies that $U_{nm}(0) = \delta_{nm} =
U_{mn}^*(0)$ and the unitarity relation implies that $\sum_p U_{pn}(t)
U_{mp}^*(t) = \delta_{nm}$ for all $t$. Now let us introduce a
``smallness'' parameter, $\epsilon$, such that
\begin{eqnarray}
U_{nm}(t) = U_{nn}(t) \delta_{nm} + \epsilon \delta U_{nm}(t)(1- \delta_{nm}).
\end{eqnarray}
Then the unitarity condition to leading order in $\epsilon$,
implies $ |U_{nn}(t)|^2 = 1$ for all t. Hence, $U_{nn}(t) = e^{i
\phi_{n}(t)}$, with $U_{nn}(0)= 1$ or $\phi_{n}(0) = 0$.  The
explicit form of the phase to this order is given by the real number
\begin{eqnarray}
\phi_{n}(t) = - \int_0^t E_n(t') dt' + 
\int_0^t i\la n(t')|{\dot n }(t') \ra~ dt'.
\end{eqnarray}
The first term is the dynamical phase and the second one gives us the
familiar Berry term. The corresponding $U(t)$ is then of the form
\begin{eqnarray}
U(t) &=& \sum_{n} e^{i \phi_n(t)} |n(t)\ra \la n(0)| \nonumber\\
& +&\epsilon
\sum_{n\not=m} |n(t)\ra \delta U_{nm}(t) \la m(0|. 
\end{eqnarray}
Thus the adiabatic theorem is presented with a consistent form.

The `smallness' parameter actually decides when to keep the off-diagonal
terms and when the approximation with diagonal ones is satisfactory. But
how small can it be? One
can obtain a non-trivial lower bound on the $\epsilon$ as follows. Let the
initial state of the system is $|\psi(0)\ra = |n(0)\ra$. The state at a later
time $t$ is $|\psi(t)\ra = U(t)|n(0)\ra$. Now the transition amplitude between
the initial and the final state is
\begin{eqnarray}
\la \psi(0)|\psi(t)\ra &=& \la n(0)|U(t)|n(0) \ra = U_{nn}(t) \la n(0|n(t)\ra 
\nonumber\\
&+& \epsilon \sum_{m\not=n} \delta U_{mn}(t) \la n(0|m(t)\ra.
\end{eqnarray}
This leads to
\begin{eqnarray}
|\la \psi(0)|\psi(t)\ra| \le |\la n(0|n(t)\ra|
+ \epsilon \sum_{m\not=n} |\delta U_{mn}(t)|.
\end{eqnarray}
The above inequality can be written as
\begin{eqnarray}
\epsilon \ge \frac{ D(|n(0),|n(t)\ra ) - D(|\psi(0)\ra, |\psi(t)\ra) } 
{\sum_{m\not=n} |\delta U_{mn}(t)| },
\end{eqnarray}
where $ D(|n(0),|n(t)\ra )$ is the 
`minimum-normed' distance functions between the initial eigenstate and 
final eigenstate, and $D(|\psi(0)\ra, |\psi(t)\ra)$ is the similar one 
between the initial state and the final state of the system \cite{akp}.
When we are within adiabatic regime $ D(|n(0),|n(t)\ra )$ and 
$D(|\psi(0)\ra, |\psi(t)\ra)$ are same, hence we have $\epsilon = 0$.
There is no lower bound on it. However, if we move away from adiabatic regime
$ D(|n(0),|n(t)\ra )$ and 
$D(|\psi(0)\ra, |\psi(t)\ra)$ differ. Then we will have a lower bound on the
`smallness' parameter.

One can also resolve the MS inconsistency using the unitary operator method.
Note that we can obtain $|{\bar n}(t)\ra = \sum_p |p(0)\ra U_{pn}^* =
U_{nn}^* |n(0)\ra +$ off-diagonal terms, and thus as before we have 
$|{\bar n}(t)\ra \approx 
\exp(i\int E_n(t') dt' ) \exp(- i \gamma_n(t)) |n(0) \ra$. Then using 
$\la n(0)|n(t)\ra \approx \exp(-i \gamma_n(t) )$ one can resolve the MS 
inconsistency.
Below we give two examples to illustrate the power of 
unitary operator method developed in this letter.

\section{Two Examples}

Here we consider two examples that will illustrate the main points.

\subsection{ The MS Example in the instantaneous representation}

 Let us consider the precession of a spin-half particle in a
magnetic field of strength proportional to $\omega_0$ and in addition it 
rotates in the $x-y$ plane with a frequency $\Omega = 2\pi /\tau$. This is
also considered by MS \cite{ms}. We work here in the instantaneous 
representation. The Hamiltonian is given by $H(t) = {\bf R}(t).\sigma$, where
$$H(t)=\left(\matrix{ \frac{\Omega}{2}(1- \cos 2\omega_0 t)  &  
e^{-i \Omega t} ( \omega_0 - i \frac{\Omega}{2}\sin 2\omega_0 t) 
\cr e^{i \Omega t} ( \omega_0 + i \frac{\Omega}{2}\sin 2\omega_0 t)  
& - \frac{\Omega}{2}(1- \cos 2\omega_0 t) } \right).$$
Its eigenvalues and eigenfunctions are $E_1(t) = \sqrt{\omega_0^2 +
\Omega^2 \sin^2 \omega_0 t} = E_0(t)$, $E_1(0) = \omega_0$; $|n_1(t)\ra = 
{\rm col}(a_1(t), b_1(t) )$,  $|n_1(0)\ra = 1/\sqrt{2}
{\rm col}(1, 1)$, with 
\begin{eqnarray}
a_1(t) &=& \sqrt{ \frac{E_0(t) + \Omega \sin^2 \omega_0t}{ 2E_0(t) }}
e^{-i(\Omega t + \theta(t) )/2 } \nonumber\\
b_1(t) &=& \sqrt{ \frac{E_0(t) - \Omega \sin^2 \omega_0t}{ 2E_0(t) }}
e^{i(\Omega t + \theta(t) )/2}.
\end{eqnarray}
And similarly, we have
 $E_2(t) = - E_0(t)$, $E_2(0) = -\omega_0$; $|n_2(t)\ra = 
{\rm col}(a_2(t), b_2(t) )$,  $|n_2(0)\ra = 1/\sqrt{2}
{\rm col}(-1, 1)$, with 
\begin{eqnarray}
a_2(t) &=& -\sqrt{ \frac{E_0(t) - \Omega \sin^2 \omega_0t}{ 2E_0(t) }}
e^{-i(\Omega t + \theta(t) )/2} \nonumber\\
b_2(t) &=& \sqrt{ \frac{E_0(t) + \Omega \sin^2 \omega_0t}{ 2E_0(t) }}
e^{i(\Omega t + \theta(t) )/2}.
\end{eqnarray}
In the above equations, $\theta(t) = \tan^{-1}[ (\Omega/2\omega_0)
 \sin 2 \omega_0t]$. 
The unitary time evolution associated with the Hamiltonian $H(t)$ is given by
$$U(t) = \left(\matrix{ \cos \omega_0 t  &  
-i e^{-i \Omega t} \sin \omega_0 t 
\cr -i e^{i \Omega t} \sin \omega_0 t
& \cos \omega_0 t } \right).$$
We need to express this in terms of the instantaneous eigenstates to isolate
the adiabatic term from the non-adiabatic pieces. We present here the diagonal
terms, being the adiabatic contributions to the evolutions:
$\la n_1(t)|U(t)|n_1(0)\ra = \frac{1}{\sqrt 2}[a_1(t) (\cos \omega_0 t -
i  e^{- i \Omega t} \sin \omega_0 t) + b_1(t) (\cos \omega_0 t -
i  e^{i \Omega t} \sin \omega_0 t)$ and 
$\la n_2(t)|U(t)|n_2(0)\ra = \frac{1}{\sqrt 2}[a_2(t) (\cos \omega_0 t +
i  e^{- i \Omega t} \sin \omega_0 t) - b_2(t) (\cos \omega_0 t +
i e^{i \Omega t} \sin \omega_0 t)$. These exact solutions 
can be evaluated explicitly and compared with the solutions based on adiabatic
approximation.

\subsection{ The Schwinger example}

This example is an elegant one originally 
due to Schwinger 
\cite{js} for describing the precession of a spin in a transverse 
time-dependent
field. Here the Hamiltonian of the system is given by $H(t) = -g \mu_0 H(
\sigma_x \sin \theta \cos \phi + \sigma_y \sin\theta \sin \phi + \sigma_z 
\cos \theta)
 = -g \mu_0 H \left(\matrix{ \cos \theta  &  
\sin \theta e^{-i \phi} \cr \sin \theta e^{i \phi} & - \cos \theta } \right)$,
where the magnetic field $H$ and $\theta$ are independent of time, and
$\phi = \omega t$. The MS model is a version of the Schwinger's NMR precession
problem described above. The instantaneous eigenvalues are $E_1 = g\mu_0H$ and
$E_2 = -g\mu_0H$. The instantaneous eigenstates are $|n_1(t)\ra =
{\rm col} ( e^{-i\phi/2} \sin \theta/2, - e^{i\phi/2} \cos \theta/2 )$ and
$|n_2(t)\ra =
{\rm col} ( e^{-i\phi/2} \cos \theta/2, e^{i\phi/2} \sin \theta/2 )$. From 
these we have $\la n_1(t)|{\dot n}_1(t) \ra = i( \omega \cos \theta)/2 =
- \la n_2(t)|{\dot n}_2(t) \ra$. Similarly, we have 
 $\la n_1(t)|{\dot n}_2(t) \ra = -i( \omega \sin \theta)/2 = 
\la n_2(t)|{\dot n}_1(t) \ra$. 
The solution to the time evolution operator
$U(t)$ in terms of the instantaneous eigenstates can be obtained as
$U(t) = \sum_{ij=1,2}^2 |n_i(t)\ra U_{ij}(t) \la n_j(0)|$ with $U_{ij}(0) =
\delta_{ij}$. The equations to be solved are 
The matrix elements of the time evolution operator obey Eq(13).
\begin{eqnarray}
iU_{11} +  i\la  n_1|{\dot n}_2 \ra U_{21}
&=& ( E_1  -  i \la  n_1|{\dot n}_1 ) \ra U_{11} \nonumber\\
iU_{21} +  i\la  n_2|{\dot n}_1 \ra U_{11}
&=& ( E_2  -  i \la  n_2|{\dot n}_2 ) \ra U_{21}.
\end{eqnarray}
A similar pair of equations for the other two. 
We can solve these equations by Laplace transforms and obtain the solutions:
The solutions are given by
\begin{eqnarray}
U_{11}(t) &=& [{\tilde E}_1 \cos {\tilde E}_1 t \nonumber\\
&-& i(g\mu_0 H + \frac{\omega}{2}
 \cos\theta ) \sin {\tilde E}_1 t ]/ {\tilde E}_1 = U_{22}^*(t) \nonumber\\
U_{21}(t) &=& i [\omega \sin\theta  \sin {\tilde E}_1 t] / 2 {\tilde E}_1 =
U_{12}(t),
\end{eqnarray}
where ${\tilde E}_1 = [(g\mu_0 H)^2 + g\mu_0 H \omega \cos\theta + 
\omega^2/4]^{1/2}$.
The unitarity of the this $U$-matrix is obeyed, as it should be. 
These are the exact solutions of the Schwinger problem. 

If we make the adiabatic approximation, we would drop the $U_{21}$ term
in Eq.(22) and obtain the following result:
\begin{eqnarray}
U_{11}(t) & \approx & \exp [ -it (g\mu_0 H + \frac{\omega}{2}
 \cos\theta ) ] = U_{22}^*(t)\nonumber\\
U_{21}(t) & \approx & i \frac{\omega}{2} \sin \theta \bigl[
\frac{ \sin t( (\omega/2) \cos \theta + g\mu_0 H)}
{( (\omega/2) \cos \theta + g\mu_0 H) }\bigr] = U_{12}(t).
\end{eqnarray}
This means that we drop terms of the order of $(\omega/2) \sin \theta$ in the
exact expressions, so that ${\tilde E}_1 \approx [g\mu_0 H + (\omega/2)
 \cos \theta]$. We thus see that the ``smallness'' parameter in the 
``adiabatic'' treatment is $\la n_1(t)|{\dot n}_2(t)\ra$. Note that the
unitarity of this $U(t)$ is verified to be obeyed consistent with the 
adiabatic approximation stated above. Also, a way to see the departure of
the result in the adiabatic approximation from the exact result is to compute
the overlap $_A\la n_1(t)|n_1(t)\ra_E$ and hence the fidelity, $F(t) =
| _A\la n_1(t)|n_1(t)\ra_E |^2$. The overlap is given by 
$_A\la n_1(t)|n_1(t)\ra_E = \exp(i {\tilde \Omega} t ) 
\bigl(\cos {\tilde E}_1 t - i({\tilde \Omega}/{\tilde E}_1) 
\sin {\tilde E}_1 t \bigr) + (\omega \sin\theta /2)^2 
(\sin {\tilde E}_1 t / {\tilde E}_1) (\sin {\tilde \Omega} t / {\tilde \Omega})
 $, where ${\tilde \Omega} =  (g\mu_0 H + (\omega/2) \cos \theta)$.
One sees at once that to the leading order in 
$(\omega \sin \theta/2)$, the fidelity is unity.


\section{Conclusions}

Adiabatic theorem though widely studied, can reveal surprising phenomena, and
sometime even apparent inconsistencies, namely, MS type or vanishing 
Berry phase type. Thus, it is of utmost importance to know where such
inconsistencies may arise.
Besides showing how to resolve these inconsistencies
associated with the adiabatic theorem (AT), we have presented here 
several novel results. We have established an equation of motion for 
the transition amplitude leading to the Berry phase, developed the 
unitary time evolution operator expressed in instantaneous basis
as a powerful tool in AT as illustrated with two examples, 
estimated the ``smallness'' parameter in the unitary operator, 
and quantified the departure of AT from the exact by means of 
``Fidelity'' expressed in terms of the respective unitary evolution 
operators. Given the current interest in the geometric phases for mixed states 
\cite{erik}, 
this work opens up the possibility of studying the Berry phase and adiabatic 
theorem for density operators and entangled quantum systems in a new 
perspective which one may like to take up in future.

{\it Note Added:} After completion of 
our work we noticed Ref.\cite{sara} which
addresses the MS inconsistency. Our present work goes beyond this 
aspect of the adiabatic theorem.

{\bf Acknowledgments:} AKP thanks K. P. Marzlin and B. C. Sanders for 
useful remarks. AKR thanks the Indo-US Workshop on Nanoscale Materials
for providing him the travel support to a visit to Puri, India which
enabled this collaboration. He also thanks the office of Naval Research for
partial support of this work.


\end{multicols}
\end{document}